\documentclass[twocolumn]{aastex7}

\newcommand{\masyr}{mas~yr$^{-1}$\,}
\newcommand{\msun}{$M_\odot$\,}

\begin{document}

\title{Mutual Orbit Alignment in Resolved  Triple Systems}


\author{Andrei Tokovinin}
\affiliation{Cerro Tololo Inter-American Observatory --- NSF's NOIRLab
Casilla 603, La Serena, Chile}
\email{andrei.tokovinin@noirlab.edu}

\begin{abstract}
A sample of 278 triple systems with outer separations under 300 au and
resolved  inner pairs  is studied,  focusing on  the mutual  alignment
between inner  and outer  orbits.  The  degree of  alignment increases
with (i) decreasing  outer separation, (ii) decreasing  ratio of outer
and  inner separations,  (iii) decreasing  mass of  the inner  primary
component,  and  (iv)  increasing  inner  mass  ratio.   There  is  no
dependence on the outer mass  ratio. The average mutual inclination is
$\sim$40\degr for  the full  sample and  $\sim$10\degr for  38 triples
with primary  components less  massive  than 1  \msun and  outer
separations below 50 au.  Inner  eccentricities in aligned triples are
smaller  compared  to  misaligned  ones.  In  another  sample  of  371
hierarchies with  known outer  orbits and inner  eclipsing subsystems,
only 22\%  show mutual  alignment within 20\degr,  while the  rest are
aligned   randomly.   These   findings  match   qualitatively  current
understanding  of the  formation  of hierarchical  systems, where  the
N-body dynamics  dominates at  large scales,  while the  accretion and
migration  shape systems  closer than  $\sim$100 au.  Fragmentation of
isolated  cores  apparently  produces approximately  aligned  low-mass
hierarchies.
\end{abstract}



\section{Introduction}
\label{sec:intro}

The  genesis  of  binary  stars and  higher-order  hierarchies  is  an
important but  still poorly understood  aspect of star  formation 
  \citep{Offner2023,Trip2021}.    The  solution  of  this problem  is
hindered by the lack of reliable statistics of high-order multiplicity
caused by  observational limitations   such as  angular resolution
  and  spectroscopic coverage.   On  the other  hand,  the theory  of
multiple-star  formation is  still  in its  infancy: diverse  physical
processes play  a substantial  role, but  their modeling  and relative
importance  is a  subject of  ongoing  effort. Joining  the two  ends,
theory and observations,  remains a challenge. In  this contribution I
address the  issue of relative  orbit alignment in triple  stars using
recent data.  In  a broader context, this work  complements studies of
relative  inclinations between  stellar  rotation axes  and orbits  of
binaries \citep{Marcussen2022}  and planets  \citep{Albrecht2022}, and
of    the   alignment    of   planetary    orbits   within    binaries
\citep{Dupuy2022,Christian2025}.

The Gaia  mission \citep{Gaia1,Gaia3} makes  fundamental contributions
in   many  areas,   including  stellar   multiplicity.   The   uniform
astrometric   and  photometric   coverage  of   the  whole   sky  with
well-defined detection  limits identifies  wide physical  binaries and
hierarchies, enabling  unbiased evaluation  of their  statistics.  The
angular resolution of the current Gaia Data Release 3 (GDR3) is on the
order of 1\arcsec, giving access  to separations above $\sim$100 au at
100 pc  distance.  The catalog  of $\sim$1 million wide  pairs derived
from   GDR3  supersedes   prior   samples  by   orders  of   magnitude
\citep{ElBadry2021}.   Extension  of  this   effort  to  triple  stars
identifies      $\sim$10,000  wide    hierarchies      within      500\,pc
\citep{Shariat2025b}.  These  authors extend  the earlier
study  by \citet{Gaiatriples}  based  on a  much  smaller but  cleaner
sample of wide triples within 100\,pc.  Their statistics largely match
the predictions  of pure  gravitational N-body dynamics,  but indicate
that at separations below 100  au the orbits become partially aligned.
The  components' masses  in wide triples do not correspond  to a random
draw from the initial mass function (IMF): they are correlated and, on
average, are larger compared to single field stars.

However, Gaia has a modest angular resolution and a limited time base.
Many new  close triples  have been discovered  during the  last decade
using  ground-based  methods,  especially among  low-mass  stars  (see
section~\ref{sec:obs}).  This  work uses  the modern speckle  data and
updates the studies of relative orbit alignment by \citet{Sterzik2002}
and \citet{Tok2017}.

The mutual inclination $\Phi$ between  the inner and outer orbits 
  (i.e.  the angle between their orbital angular momentum vectors) in
a triple system can be computed from the known orbital elements. 
  Aligned, orthogonal, and anti-aligned orbits correspond to $\Phi=0$,
  $\Phi = \pi/2$, and $\Phi=  \pi$, respectively.   The  elements
  of the inner and outer orbits  with correct node identification are
available only  for a small  sample.  However, a  powerful statistical
measure of  mutual alignment is  the correlation between  the observed
sense of motion,  i.e. the numbers of co-  and counterrotating triples
$n_+$ and $n_{-}$ \citep{Worley1967,Sterzik2002}. The sign correlation
$C$ is defined as
\begin{equation}
C = (n_+ - n_{-})/(n_+ + n_{-}) = 1 - 2 \langle \Phi \rangle/\pi .
\label{eq:C}
\end{equation}
If  co- and  counterrotating triples  are assigned  $S=1$ and  $S=-1$,
respectively, the sign correlation is simply the average value of $S$.
The relation between  $C$ and the average  mutual inclination $\langle
\Phi  \rangle$  follows  from simple  geometric  considerations.   For
randomly aligned orbits,  $C = 0$ and $\langle \Phi  \rangle = \pi/2$.
This method    does not
require knowledge  of two orbits and  nodes, hence is applicable  to a
much larger sample.  However, $C$ has only a statistical meaning.  All
aligned systems  corotate, but the  inverse is  not true: half  of the
randomly aligned systems also corotate, so the significance of $S$ for
individual triples  is not high.   The motion sense is  poorly defined
for orbits seen edge-on.  Excluding  such systems (as done here) gives
more robust  results, but the  relation between $C$ and  $\langle \Phi
\rangle$ becomes  only approximate  because the assumption  of uniform
orientation of orbits relative to the observer is no longer accurate.

Here  I study  resolved triple  systems with  outer separations  below
$\sim$300 au to explore the  transition from  misalignment at large
separations to   prevailing alignment of  more compact hierarchies.
The  increased size  of  the  modern sample  allows  us  to probe  the
dependence  of  the  mutual  alignment on  the  separation  and  other
parameters  such  as  mass  and  mass  ratio.   Section~\ref{sec:data}
presents the observational data, the  sample, and the determination of
relative  motions.   Statistics of  this  sample  are the  subject  of
Section~\ref{sec:stat}, where  the additional samples of  2+2 resolved
quadruples   and  triples   with   inner   eclipsing  subsystems   are
characterized.  Section~\ref{sec:disc}  compares findings of this
  study with theory and simulations.

\section{Data}
\label{sec:data}

\subsection{Observations}
\label{sec:obs}

In  recent  years, adaptive optics and  speckle interferometry have
opened a new window  into  the low-mass hierarchical systems among nearby
stars.  These  objects are  found during  multiplicity surveys  in the
field  \citep[e.g.][]{Leinert1997,Jnn2014b,Winters2019, Clark2024}  and
 in young   associations  \citep{USco}.   New  hierarchies   are  also
discovered  serendipitously in  high-resolution observations  of known
visual  binaries,  revealing one  of  their  components as  a  tight
pair. The growing number of  known triples and higher-order systems is
reflected in  the Multiple Star  Catalog (MSC) \citep{MSC}.   Its last
(December    2023)    public   update    available    online\footnote{
  \url{http://vizier.u-strasbg.fr/viz-bin/VizieR-4?-source=J/ApJS/235/6}
  and  \url{https://www.ctio.noirlab.edu/~atokovin/stars/} }  contains
$\sim$4000 systems. Each hierarchy is identified by its MSC code based
on the  J2000 position,  similar to (and,  mostly, coincident with)  such
codes in the Washington Double Star Catalog (WDS) \citep{WDS}.

To determine  relative motions and orbital  configurations of resolved
hierarchies, their  discovery should be followed  by monitoring during
several years or decades.  Such a long-term program is being conducted
at the 4.1 m Southern Astrophysical Research (SOAR) telescope in Chile
using speckle  interferometry.  The instrument and  data reduction are
covered   in  \citep{TMH10,HRCam}.    The  speckle   measurements  and
resulting orbits are published in a series of papers \citep[the latest
  is][]{Tok2024}  and  used  for  the  study  of  individual  resolved
hierarchies
\citep{TL2017,Dancingtwins,TL2020,Trip2021,Trip2023,Trip2025,chiron11}.
These data are incorporated in the present work.

\begin{figure}
\epsscale{1.0}
\plotone{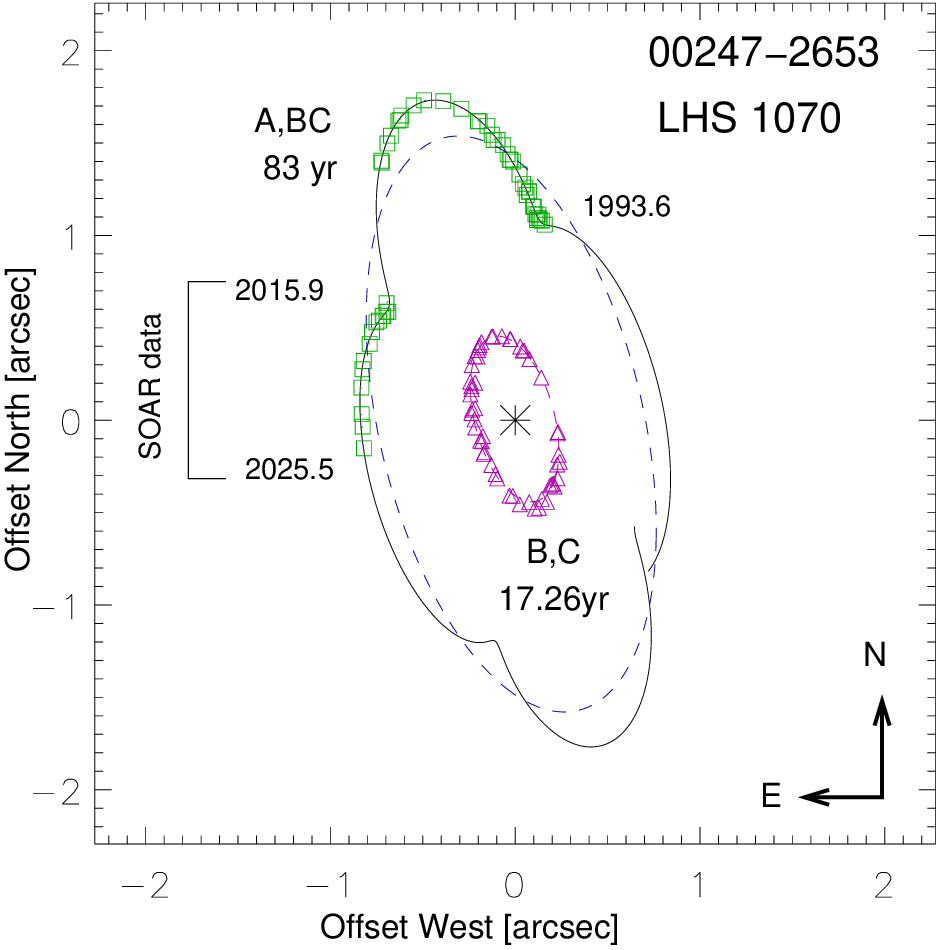}
\caption{Observed  orbital  motions  in  the  low-mass  triple  system
  00247$-$2653 (LHS  1070) are  approximated by two  Keplerian orbits,
   highlighting the decade of  monitoring at SOAR.  The positions
  of the  inner pair  B,C are  plotted by  magenta triangles,  and the
  positions of A,B  by green squares.  The solid line  shows the outer
  orbit with wobble, the dashed line  is the outer orbit of the center
  of mass BC around A.
\label{fig:LHS1070}
}
\end{figure}

To illustrate the speckle monitoring at SOAR, Figure~\ref{fig:LHS1070}
plots the  data for  00247$-$2653 (LHS  1070). This  remarkable triple
system bordering the  stellar and substellar regimes  (masses of 0.15,
0.07,  0.07  \msun)  has  been discovered  by  \citet{Leinert1997}  in
1993.6.  Owing to  the strong interaction between the  inner and outer
subsystems, the  approximation of  motion by  two Keplerian  orbits is
rather crude \citep{Xia2019}.  Nevertheless,  the two-orbit fit to the
data up to 2025.5 yields the inner and outer periods of 17.26$\pm$0.01
and  83.06$\pm$1.71 years,  respectively,  and  the eccentricities  of
$e_{\rm in} = 0.0170 \pm 0.0008$  and $e_{\rm out} = 0.036 \pm 0.007$.
The mutual inclination  is 2\fdg0$\pm$0\fdg4, and the  period ratio is
4.81$\pm$0.10,  near   the  limit  of  the   dynamical  stability  and
suggestive of a 1:5 mean  motion resonance.  Although three decades of
monitoring cover only a fraction of  the outer orbit, its character is
already well constrained.  The low-mass (M4.5V) triple 10367+1522 (DAE
3) has  a similar architecture  with periods of  8.6 and 120  yr, small
eccentricities, and aligned orbits.  Another interesting triple system
DENIS   J020529.0$-$115925   consisting    of   three   brown   dwarfs
\citep{Bouy2005}  with an  outer separation  of 8  au still  lacks 
data  on its orbital motion.

The  detailed  view  of  the orbital  dynamics  illustrated  above  is
possible  only  for a  few  well-studied  triples.   For  the
recently  discovered systems,  the available  measurements cover  only
small orbital arcs, especially in the slowly moving outer pairs.

\subsection{Sample}
\label{sec:sam}

The sample of resolved hierarchies is  based on the current version of
the  MSC.   This  is   a  heterogeneous  collection  of  observations,
incomplete   even  within   the  100\,pc   volume  \citep{GKM}.    The
observational limits in resolution and contrast leave a strong imprint
on the  statistics of  separations and mass  ratios of  those systems.
However,  these  limitations  should  not  affect  the  observed  {\em
  relative motions}.   This study focuses  on the analysis  of orbital
motions,  recognizing  that  other   parameters  in  this  sample  are
selection-dependent.

Masses  in the  MSC are  derived  from the  absolute magnitudes  using
standard relations for dwarfs \citep{Pecaut2013}.  The period $P^*$ is
estimated from  the projected separation  (its median is close  to the
semimajor  axis), unless  the  true  period $P$  is  known.   The
  statistical relation between projected separation and semimajor axis
  is covered in several papers, e.g.  in the Appendix of \citet{Orion}
  and in \citet{Makarov2025}.    In this, study  I select hierarchies
with  outer  periods  shorter  than $10^{6}$  days  (2.7  kyr),  which
corresponds to  separations below $\sim$300  au.  This cut  is imposed
because wider hierarchies in the  field are already well characterized
by Gaia, and  here I want to explore in  greater detail the transition
from random orbit orientation to alignment.

To  allow determination  of  the inner  orbit  orientation, the  inner
subsystem  should have  a known  astrometric  or visual  orbit or,  at
least,  be  directly  resolved.   I eliminate  a  few  resolved  inner
subsystems with  periods shorter  than 100  days and  resolved triples
where  the  inner  pair  contains an  even  closer  (e.g.   eclipsing)
subsystems, but retain hierarchies where the resolved triple has wider
companion(s). The reason is  to study only triples within limited
  range of inner  and outer periods.  The resolved  quadruples of 2+2
type (two  close pairs) are  also excluded from this  sample and
discussed     separately     in     section~\ref{sec:quadr},     while
section~\ref{sec:visecl}  explores the  alignment  of inner  eclipsing
pairs  in triples with  resolved outer components.

 The original selection of 355  candidates from the MSC is reduced
  to 278 systems where the orbits or the directions of relative motion
  could   be  determined   in   both  outer   and   inner  pairs.    
Table~\ref{tab:sample} in  section~\ref{sec:tables}  lists these
hierarchies and their main parameters.  The inner and outer orbits are
known in 176  and 76 triples, respectively.  Although  some orbits are
tentative  or preliminary  (especially  with long  periods),  the
  motion direction  is defined reliably.    The remaining  102 inner
and  202 outer  subsystems have  known direction  of relative  motion,
discussed in the following section.

The  present   sample  partially   overlaps  with  my   earlier  study
\citep{Tok2017}, but  differs in several  ways. The previous  work did
not restrict  the outer periods  and included both the  2+2 quadruples
(considered as two  triples) and the resolved  triples with additional
inner subsystems. The restrictions  imposed here became feasible owing
to the  growing content of the  MSC.  Most objects in  this sample are
relatively recent discoveries from  high-resolution imaging surveys of
low-mass stars;  the new sample  includes triples without  known inner
orbits not  considered in the  earlier study.

\subsection{Relative Motions}
\label{sec:meth}

In  this study,  the  published measurements  retrieved  from the  WDS
database  \citep{WDS} are  combined with  the SOAR  data.  The  median
parallax is 15\,mas, so for half of the sample the outer separation of
300 au corresponds  to an angular separation of  $>$4\farcs5, and GDR3
can contain  astrometric solutions for both sources.  However,
the  proper motions  of unresolved  inner  pairs are  biased by  their
orbital  motions  (except those  with  Gaia  astrometric orbits)  and,
lacking good models  of these motions for the mean  epoch of 2016.0, I
cannot compute  and subtract this  bias to  isolate the motion  of the
outer   pair,   as  has   been   done   for  resolved   Gaia   triples
\citep{Gaiatriples}.  Here I  rely only on the  relative positions and
orbits.

\begin{figure}
\epsscale{1.0}
\plotone{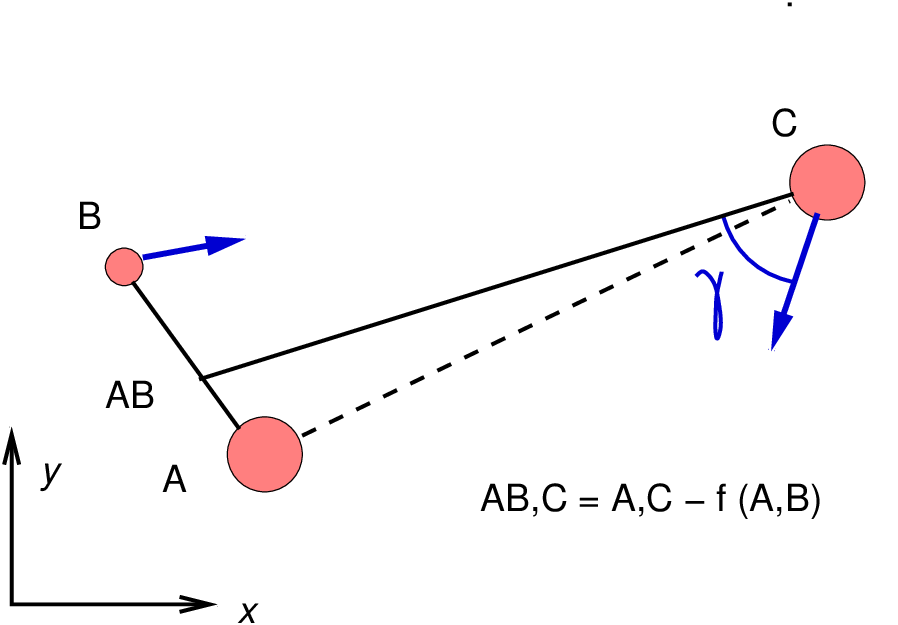}
\caption{Relative  motions  in  a  triple system  of  hierarchy  AB--C
  (section~\ref{sec:meth}).  AB  is the  center of  mass of  the inner
  subsystem A,B. Blue arrows indicate the motion directions.   The
    angle  between  the radius-vector  joing  the  components and  the
    motion direction is $\gamma$.
\label{fig:triple}
}
\end{figure}

The relative astrometry  of resolved triples refers  to the individual
stars: positions of the secondary  component B relative to the primary
A in  the inner subsystem  and of the tertiary  star C relative  to A.
The positions  are measured as a  function of time, as  illustrated in
Figure~\ref{fig:triple}.  The  polar coordinates $(\rho,  \theta)$ are
converted into Cartesian coordinates $x,y$,  where  $x$ points
to  the north  and $y$  to the  east.  The  observed trajectories  are
approximated by linear functions of time, e.g.
\begin{equation}
x(t) = \rho(t) \cos \theta(t) \approx x_0 + \dot{x} (t - t_0),
\label{eq:x(t)}
\end{equation}
where $t_0$ is  the mean time of observations. The  position vector at
$t_0$ is  $(x_0, y_0)$,  and the relative  speed vector  is $(\dot{x},
\dot{y})$.   The   angle  $\gamma$  between  these   vectors  contains
information  on   the  direction  of   the  relative  motion   (it  is
counterclockwise, also  called direct, for  $0 < \gamma  < 180^\circ$)
and  is statistically  related to  the distribution  of eccentricities
\citep{TokKiy2016}.

The motion  of the  outer pair should  be in  reference   to the
center of mass of  the inner subsystem, denoted as AB.   To do so, the
measured  relative  positions  of  A,C are  corrected  by  adding  the
positions of A,B scaled by the wobble  factor $| f | = q_{\rm in}/(1 +
q_{\rm  in})$,  depending  on  the  inner mass  ratio  $q_{\rm  in}  =
m_B/m_A$.  This factor  is negative when the subsystem  belongs to the
primary, as shown in Figure~\ref{fig:triple}, and positive when it
is in the secondary (A--BC hierarchy).  When the inner orbit is known,
the  positions are  computed from  its elements;  otherwise, from  the
linear  model (\ref{eq:x(t)}).  If the  outer  arc is  measured
during several  inner periods,  the inner motion  does not  affect the
linear fit of the outer pair.


\begin{figure}
\epsscale{1.0}
\plotone{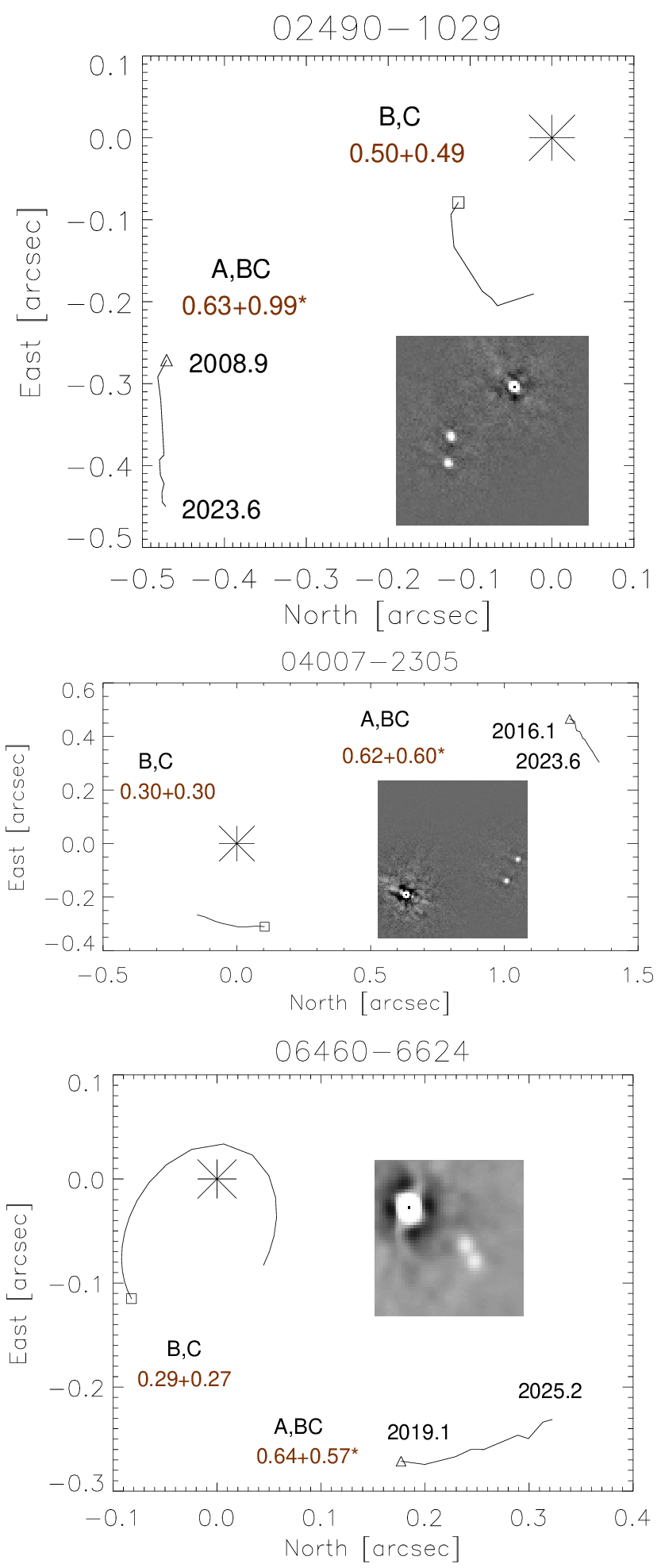}
\caption{Observed relative motions in three selected low-mass triples.
  The  coordinate origin  is  marked by  a  large asterisk.   Relative
  positions of the  inner and outer pairs are plotted  by solid lines,
  with the  times of the  first and last observations  indicated.  The
  brown numbers are component's masses.   The outer positions refer to
  the center  of mass of the  inner pair.  The first  positions of the
  inner  and  outer pairs  are  marked  by  the square  and  triangle,
  respectively.  The inserts show the  speckle images of these triples
  from the latest observations at  SOAR. In 06460$-$6624, the observed
  part of the 11.7 yr inner orbit of B,C is plotted; note the opposite
  motion sense of the inner and outer pairs in this triple.
\label{fig:motions}
}
\end{figure}

Figure~\ref{fig:motions} illustrates three  low-mass triples from this
sample.  The first  one, 02490$-$1029, was discovered  in 2008 by
\citet{Jnn2014b}.   Currently its  configuration is  non-hierarchical,
with  the inner  and  outer separations  of  0\farcs19 and  0\farcs58,
respectively.  Both  pairs move  counterclockwise.  The  second triple
04007$-$2305  (GJ 3260)  has  been  discovered at  SOAR  in 2016;  its
subsystems move clockwise.  In both triples the observed arcs are much
larger than the measurement errors, allowing reliable determination of
the  relative  speed.   The   third  example  illustrates  the  triple
06460$-$6624   (discovered   at   SOAR in 2019)  with   counterrotating
subsystems and known inner orbit.

Approximation  of the  observed arcs  by linear  functions of  time is
quite  crude,  but sufficient  for  the  purpose  of this  study.   To
simplify,  no weights  are  used  in the  linear  fits when  combining
measurements  from  different  sources.    The  errors  of  the  speed
components $\dot{x}$ and  $\dot{y}$ are determined from  these fits if
three or more  measurements are available;   for two measurements,
  the unknown  errors are  set to  zero.  The  speed of  the relative
motion $\mu = \sqrt{ \dot{x}^2 + \dot{y}^2 }$ can be normalized by the
nominal speed, $\mu' = \mu/\mu^*$ \citep{TokKiy2016}, where
\begin{equation}
\mu^* = 2 \pi \varpi^{1.5} \rho^{-0.5} M^{0.5} .
\label{eq:mustar}
\end{equation}
In  this formula,  $\rho  =  \sqrt{ x_0^2  +  y_0^2}$  is the  angular
separation, $\varpi$ is the parallax, and $M$ is the mass sum in solar
units. The  parameter $\mu^*$ corresponds  to the speed in  a circular
face-on orbit  with semimajor axis  $\rho$.  For bound pairs,  $\mu' <
\sqrt{2}$ (all  outer pairs in  this sample  have $\mu' <  1.5$).  The
parameters $\mu'$  and $\gamma$  are computed  from the  fitted linear
models.  The   error  of   $\gamma$  is  estimated   approximately  as
$\sigma_\gamma  \approx 57\fdg3  \; \sigma_\mu  /\mu$, assuming  equal
speed errors in the radial and tangential directions.

In  addition  to  the  linear  models,  the  observed  motion  can  be
characterized by  the areal constant  $A_z = 0.5  \rho^2 \dot{\theta}$
determined  from   the  observations.   It  is   proportional  to  the
projection of the  total angular momentum of the orbit  on the line of
sight.   The full  areal constant  $A$  (area of  the orbital  ellipse
divided  by the  period)  is related  to the  semimajor  axis $a$  (in
angular units), period $P$, and eccentricity $e$ as
\begin{equation}
A = \pi a^2 \sqrt{1 - e^2} /P = \pi \sqrt{1 - e^2}  a^{0.5} \varpi^{1.5} M^{-0.5} , 
\label{eq:A}
\end{equation}
where  the last  expression is  obtained  from the  third Kepler  law.
Obviously, $A_z = A \cos i$. If the semimajor axis $a$ were known, the
ratio $A_z/A$ would  inform us on the orbital  inclination $i$.  For crude
estimates, the factor  $\sqrt{1 - e^2}$ can be neglected  (it is above
0.87 for $e<0.5$), and $a$ can  be replaced by the observed separation
$\rho$.      For     the      triple     02490$-$1029     shown     in
Figure~\ref{fig:motions},  the  inner pair  has  $A_z/A  =0.88$ ---  a
strong indication  of a face-on orbit.   For the outer pair,  
$A_z/A = 1.9$, which means that  the current separation $\rho$ is less
than the true  semimajor axis $a$, thus underestimating  $A$.  So, the
apparently non-hierarchical  configuration of this triple  is a result
of projection.  For the  counterrotating triple 06460$-$6624, $A_z/A =
1.4$, also  indicating that  $a$ is  larger than  $\rho$ in  the outer
orbit.  In  these ``trapezia''  triples, the  outer periods  $P^*$ are
underestimated  because the  projected  separations  happen to  be
smaller than the true semimajor axes.

We can invert the relation (\ref{eq:A}) to estimate the lower limit of
the semimajor  axis $a$ in  angular units from the  measured parameter
$A_z$, without using the separation:
\begin{equation}
a = \frac{A_z^2 M}{\pi^2 \varpi^3 (1 - e^2) \cos^2 i}  > A_z^2 M \pi^{-2}
\varpi^{-3} = a_{\rm min} .
\label{eq:a}
\end{equation}
For random orbit  orientation, $\cos i$ is  distributed uniformly, the
median of  $\cos^2 i$ is  0.25, and  $a > 4  a_{\rm min}$ is  a better
statistical   estimate.     For   the    apparently   non-hierarchical
configurations of  02490$-$1029 and 06460$-$6624, the  $4 a_{\rm min}$
axes of the outer orbits are 7\farcs8 and 2\farcs3, respectively, much
larger than their current outer separations of 0\farcs6 and 0\farcs4.

\subsection{Data tables}
\label{sec:tables}

Examples of  the tables are  described here,  and the full  tables are
available   as  data   files.   Table~\ref{tab:sample}   lists  global
parameters of the hierarchical systems in our sample. The first column
gives the  MSC code  based on the  J2000 position; it  can be  used to
access complementary information contained in  the MSC.  The masses of
the inner pair $m_1$ and $m_2$ and of the tertiary component $m_3$ are
given in  the next three  columns, followed by the  parallax $\varpi$.
Then the decimal logarithm of the period in days and the semimajor axis or
projected  separation  in  au  are  given  for  the  inner  and  outer
subsystems.  The solution types indicate whether an orbit (2) or only
a  linear model  (1) are  available.  The sign  correlation $S$  takes
values  of +1  and $-$1  for corotating  and counterrotating  triples,
respectively, and  zero for systems  excluded from the  statistics for
the reasons given in section~\ref{sec:sign}.  The last column contains
the separation of the outer (fourth) companion in au, if present.

\begin{deluxetable*}{l rrrr   rr  rr  cc c l }[ht]     
\tabletypesize{\scriptsize}                    
\tablecaption{Sample Summary  (Fragment)                                                                                                                           
\label{tab:sample} } 
\tablewidth{0pt}          
              
\tablehead{ \colhead{MSC} & 
\colhead{$m_1$} &  
\colhead{$m_2$} &  
\colhead{$m_3$} &  
\colhead{$\varpi$} &  
\colhead{$\log P_{\rm in}$} & 
\colhead{$a_{\rm in}$} & 
\colhead{$\log P_{\rm out}$} & 
\colhead{$a_{\rm out}$} & 
\multicolumn{2}{c}{Solution} & 
\colhead{$S$ } &
\colhead{$a_{\rm ext}$} \\
\colhead{(J2000)} &
\multicolumn{3}{c}{(\msun)} &
\colhead{(mas)} & 
\colhead{(d)} &
\colhead{(au)} & 
\colhead{(d)} &
\colhead{(au)} & 
\colhead{In} & 
\colhead{Out} & 
& 
\colhead{(au)} 
}
\startdata
00164$-$2235 & 1.40 &0.99 &0.72 & 9.84  & 4.02 & 12.6  & 5.39 &111.8 &2 &1  & 1 &    0 \\
00174$+$0853 & 1.28 &1.13 &1.26 & 14.47 & 4.11 & 14.5  & 5.93 &272.6 &2 &1  & 0 &    0 \\
00247$-$2653 & 0.08 &0.08 &0.11 &129.30 & 3.80 &  3.7  & 4.48 & 12.4 &2 &2  & 1 &    0 \\
00304$-$6236 & 0.54 &0.49 &0.23 &22.34  & 4.05 &  9.9  & 5.96 &198.1 &2 &1  & 0 &    0 \\
00325$+$6714 & 0.41 &0.16 &0.31 &101.09 & 3.75 &  5.1  & 4.91 & 35.1 &2 &2  & 1 &    0 \\
00329$-$0434 & 0.21 &0.10 &0.12 & 52.85 & 3.72 &  4.0  & 4.65 & 18.5 &2 &1  & 0 &    0 \\
01158$-$6853 & 1.38 &0.30 &0.88 & 47.65 & 3.47 &  4.8  & 5.64 &154.5 &2 &2  & 1 & 6692 \\
\enddata 
\end{deluxetable*}

\begin{deluxetable*}{l c r r r r r r r l }
\tabletypesize{\scriptsize}                    
\tablecaption{Orbital Elements  (Fragment)                                                                                                                           
\label{tab:orb} } 
\tablewidth{0pt}          
              
\tablehead{ 
\colhead{MSC} & 
\colhead{In/Out}  &
\colhead{$P$}  &
\colhead{$T$}  &
\colhead{$e$}  &
\colhead{$a$}  &
\colhead{$\Omega$}  &
\colhead{$\omega$}  &
\colhead{$i$}  &
\colhead{Reference\tablenotemark{a}}  \\
\colhead{(J2000)} &  &
\colhead{(yr)} &
\colhead{(yr)} & &
\colhead{($''$)} &
\colhead{(deg)} &
\colhead{(deg)}  &
\colhead{(deg)}  &
}
\startdata
00164$-$2235 &I    &28.834 & 2000.281 & 0.674 & 0.144&  61.1 & 38.9& 137.5 & Tok2023a \\
00174$+$0853 &I    &35.660 & 1950.700 & 0.002 & 0.189& 124.1 &282.0&  95.4 & Hrt2010a \\
00247$-$2653 &I    &17.260 & 2006.445 & 0.017 & 0.460&  14.9 &202.5&  62.1 & New \\
00247$-$2653 &O    &83.060 & 2070.117 & 0.036 & 1.592&  13.2 &285.2&  63.3 & New \\
00304$-$6236 &I    &30.000 & 2016.880 & 0.430 & 0.202&  95.8 & 41.1& 132.2 & New \\
\enddata 
\tablenotetext{a}{Orbit references are provided at
  \url{https://crf.usno.navy.mil/data_products/WDS/orb6/wdsref.html} and in the data file. }
\end{deluxetable*}

\begin{deluxetable*}{l c  r rr rr rr rr }
\tabletypesize{\scriptsize}                    
\tablecaption{Linear Elements  (Fragment)                                                                                                                           
\label{tab:lin} } 
\tablewidth{0pt}          
              
\tablehead{ 
\colhead{MSC} & 
\colhead{In/Out}  &
\colhead{$t_0$}  &
\colhead{$x_0$}  &
\colhead{$y_0$}  &
\colhead{$\dot{x}$}  &
\colhead{$\dot{y}$}  &
\colhead{$\sigma_{\dot{x}}$}  &
\colhead{$\sigma_{\dot{y}}$}  &
\colhead{$\gamma$}  &
\colhead{$\sigma_\gamma$}  \\
\colhead{(J2000)} &   &
\colhead{(yr)} &
\colhead{($''$)} &
\colhead{($''$)} &
\multicolumn{2}{c}{(\masyr)} &
\multicolumn{2}{c}{(\masyr)} &
\colhead{(deg)} &
\colhead{(deg)}  
}
\startdata
00164$-$2235 &O &2020.17 &$-$0.244 &$-$0.997 & $-$7.4 &  3.2    & 0.4 &  0.9 &260.0 & 7.0 \\
00174$+$0853 &O &1933.79 &$-$2.464 &$-$3.510 &    1.4 &  5.1    & 0.4 &  0.8 &199.7 & 9.3 \\
00304$-$6236 &O &2006.19 &$-$2.219 &$-$3.703 & $-$8.9 &$-$14.3  & 3.1 &  1.0 &359.0 &11.0 \\
00329$-$0434 &O &2022.56 &$-$0.927 &$-$0.243 &$-$31.4 &$-$16.8  & 2.5 &  1.3 & 13.5 & 4.5 \\
00541$+$6626 &O &2010.19 &$-$0.295 &   0.828 &    4.4 &  2.5    & 0.5 &  0.4 &279.5 & 7.3 \\
01036$-$5516 &O &2009.48 &   1.566 &$-$0.752 &    1.4 & $-$7.9  & 2.2 &  5.1 &305.9 &39.4 \\
\enddata 
\end{deluxetable*}

The   orbital   elements   used   in  this   study   are   listed   in
Table~\ref{tab:orb}. Each pair is identified by the MSC code and a tag
I  (inner) or  O  (outer). The  7 elements  are  provided in  standard
notation ($P$ --- orbital period, $T$ --- epoch of periastron, $e$ ---
eccentricity, $a$  --- semimajor  axis, $\Omega$  --- position  angle of
ascending  node,  $\omega$  ---   longitude  of  periastron,  $i$  ---
inclination). The last column gives  reference codes to these orbits
borrowed from the 6th orbit catalog \citep{VB6} and complemented where
necessary; ``New''  indicates unpublished orbits computed  or modified
by the author.

Table~\ref{tab:lin} lists linear models for pairs without known
orbits. Its first two columns are same as in Table~\ref{tab:orb}. Then
follow the mean time $t_0$, coordinates $x_0$ and $y_0$ at this time,
the velocity components $\dot{x}$ and $\dot{y}$, their formal errors,
the angle $\gamma$, and its error. 

\section{Statistics}
\label{sec:stat}

\subsection{Periods and Masses}

\begin{figure}
\epsscale{1.0}
\plotone{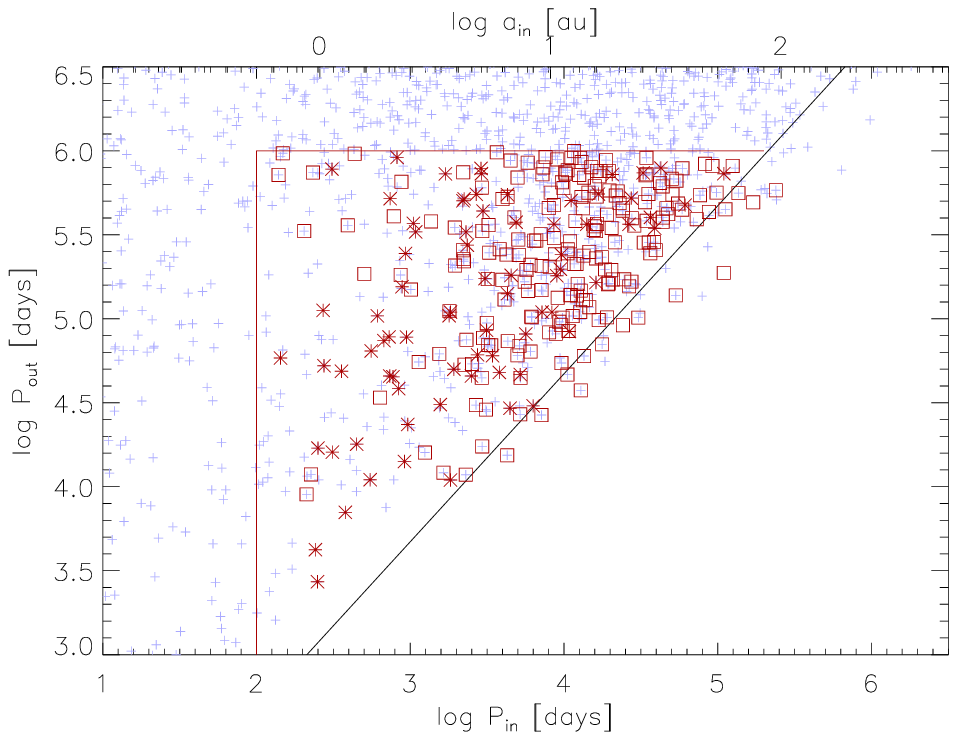}
\plotone{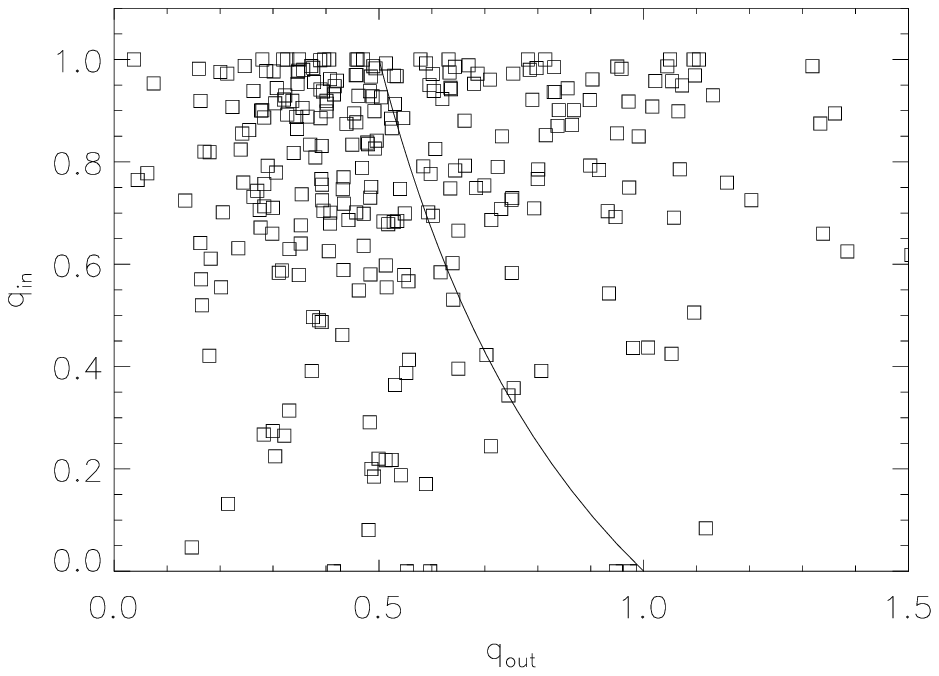}
\caption{Top: periods  of the inner  and outer subsystems  in multiple
  stars.  Small  blue crosses  denote hierarchies within  200\,pc with
  primary  masses less  than 1.5  \msun.  Large  red symbols  mark our
  sample  of  triples  with  (asterisks) or  without  (squares)  known
  orbits.   The   thin red  line  denotes the  imposed limits  on
  periods, and  the   thick solid diagonal line  plots the
  nominal dynamical stability criterion $P_{\rm out}/P_{\rm in} > 4.7$
  \citep{Mardling2001}.  Bottom: comparison between mass ratios in the
  inner and outer subsystems of  our sample (squares).  In 106 triples
  located to the  right of the line, the tertiary  is the most massive
  star.
\label{fig:q}
}
\end{figure}

Figure~\ref{fig:q}, top,  plots the  inner and  outer periods  in this
sample  (red squares and  asterisks) in  comparison with  other
known hierarchies (pale plus crosses).  The combination of the imposed
period cuts and the requirement to resolve the inner subsystem  in
  our sample results in clustering  of the points above the dynamical
stability limit  $P_{\rm out}/P_{\rm in} >  4.7$ \citep{Mardling2001}.
Points  below this  line have  underestimated outer  periods owing  to
projection, as explained above  in section~\ref{sec:meth}. 

The median masses of the  inner primary and secondary components $m_1$
and $m_2$ are  0.88 and 0.70 \msun, respectively, and  the median mass
of the  tertiary components $m_3$  is 0.80  \msun; in 106  triples the
tertiary is the most  massive star.  Figure~\ref{fig:q} (bottom) plots
the inner  mass ratio  $q_{\rm in}  = m_2/m_1$  vs.  outer  mass ratio
$q_{\rm out} = m_3/(m_1 + m_2)$.  The preference of large $q_{\rm in}$
could  be a  selection effect  (pairs  with small  mass ratios  escape
detection  in  the  imaging  surveys), although  it  agrees  with  the
simulations  by  \citet{Rohde2021}  discussed below    in  section
  ~\ref{sec:theory}.   The  apparent  concentration  of  points  near
$(0.5,1)$ corresponds to three  stars of comparable masses (triplets),
and points near  (1,1) are double twins  with  the inner and outer
  mass    ratios    close   to    one  (the    triple   shown    in
Figure~\ref{fig:LHS1070} is a double twin).

\subsection{Orbital Inclinations}
\label{sec:align}

\begin{figure}
\epsscale{1.0}
\plotone{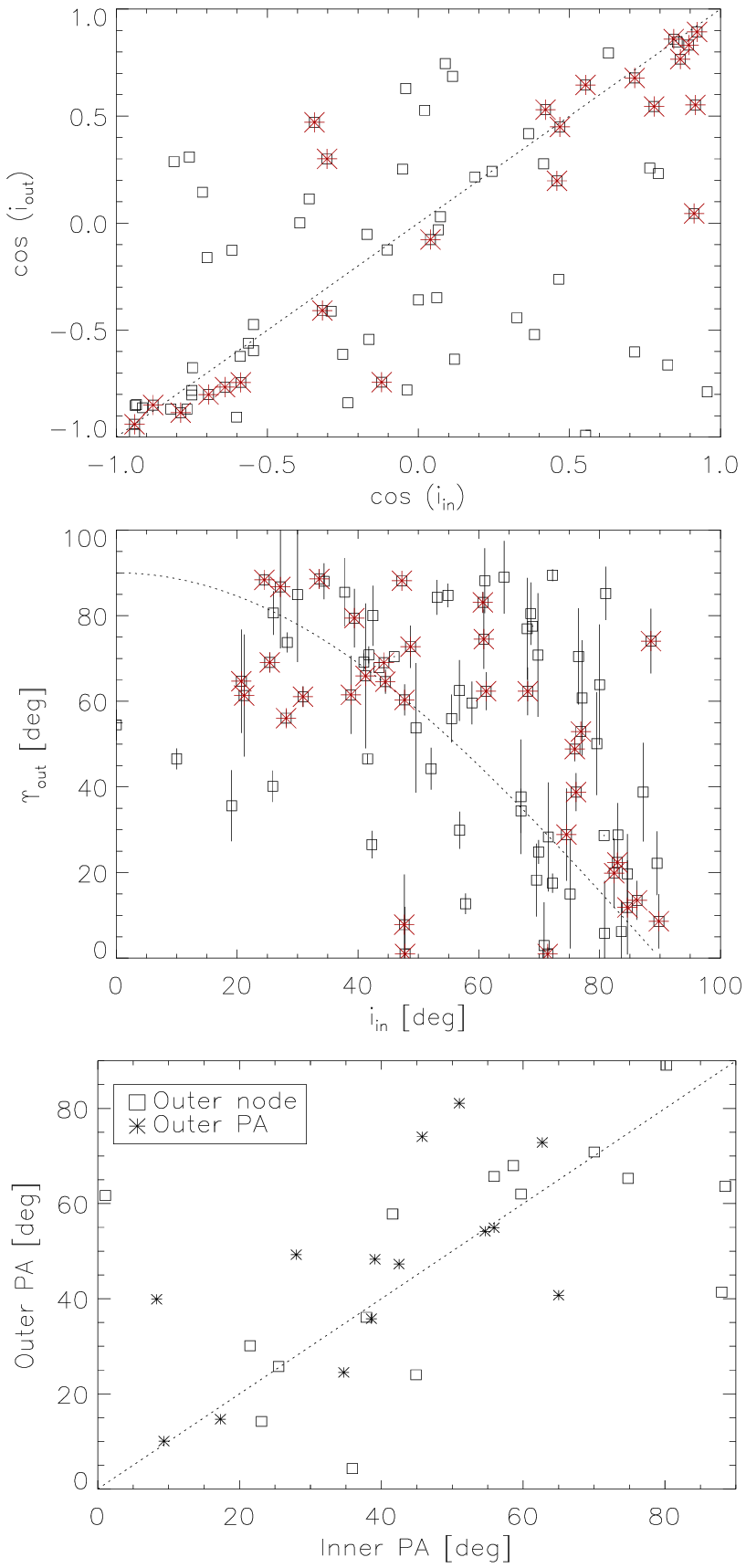}
\caption{Top: comparison  of the  cosines of  inner and  outer orbital
  inclinations for 76 triples where both orbits are known. Middle: the
  angle $\gamma$ in the outer pair  vs. inclination of the inner orbit
  for 85  triples; the dotted line  is $90^\circ \cos i_{\rm  in}$. In
  both plots, red asterisks mark systems with inner primary mass below
  0.8 \msun.  Bottom: position  angle of the  outer pair  vs. position
  angle of  the inner pair for  edge-on inner orbits; both  angles are
  folded in the $(0,90\degr)$ interval. Squares and asterisks plot the
  nodal angles  of 16 pairs with  known outer orbits and  the position
  angles of 14 outer companions with linear models, respectively.
\label{fig:incl}
}
\end{figure}

The most complete information on the  motion is available for the
76 triples where both inner and  outer orbits are known.  The relative
angle $\Phi$  between the  orbital angular  momentum  vectors is
computed as
\begin{equation}
\cos \Phi = \cos i_{\rm in} \cos i_{\rm out} + \sin i_{\rm in} \sin
i_{\rm out} \cos(\Omega_{\rm in} - \Omega_{\rm out}) ,
\label{eq:phi}
\end{equation}
where  $i$  are the inclinations and $\Omega$  the position angles
  of  ascending nodes.  However, the  true ascending  nodes are  not
known for  most of  these triples,  so each pair  of orbits  gives two
possible values of  $\Phi$ corresponding to a  180\degr ~difference in
$\Omega_{\rm in} - \Omega_{\rm out}$.  For example, if both orbits are
seen edge-on and their orientation on  the sky is equal, $\Phi$ can be
either 0  or 180\degr, and  we cannot distinguish between  the aligned
and counteraligned cases.  To  circumvent the problem, two alternative
values of $\Phi$ are computed for  each triple with known orbits.  The
joint distribution  of these angles indicates  partial alignment, 
  in    agreement   with    my   previous    result   \citep[Figure~4
  in][]{Tok2017}.

The similarity  between orbital inclinations  is a necessary  (but not
sufficient) condition for aligned orbits.  Figure~\ref{fig:incl} (top)
compares the cosines of inclinations for triples with two known orbits;
it  indicates    prevailing  alignment, with  a  few  outliers.   For
example, 20396+0458 (HIP  101955) is a triple with  reliably known and
non-coplanar ($\Phi = 64\fdg8  \pm 1\fdg4$) orbits \citep{TL2017}; its
inner  orbit  has  a   substantial  eccentricity  of  $e=0.617$.   The
correlation  coefficient between  the  cosines  is 0.60$\pm$0.09  (its
error is  evaluated by  bootstrap).  If  the sample  is split  into 23
systems with inner  primaries less massive than 0.8 \msun  and 53 more
massive  triples,  the   correlation  coefficients  between  the
inclination cosines are 0.88$\pm$0.05 and 0.44$\pm$0.13, respectively.
This indicates a stronger orbit alignment at lower mass.

The middle  panel of Figure~\ref{fig:incl} compares  the inclination of
the inner  orbit with  the angle  $\gamma'$ in the  outer pair  for 85
triples  with such  data  and $\sigma_\gamma  <  20\degr$. The  angles
$\gamma' = \arcsin ( |\sin \gamma |)$ and the inclinations are folded into
the $(0,90\degr)$ interval.  If the orbits are aligned, radial motions
are expected  for the  highly inclined orbits  with $i_{\rm  in}$ near
$90^\circ$.  Indeed,  such a  trend is  seen,  despite several
outliers. 

Finally, the lower plot in Figure~\ref{fig:incl} compares the position
angles  on the  sky for  30 systems  with approximately  edge-on inner
orbits ($i_{\rm in}$  from 80\degr to 100\degr).  For  16 systems with
both orbits known, the position angles of the ascending nodes $\Omega$
are folded in the $(0,90\degr)$  interval as indicated above.  For the
additional 14 triples without known  outer orbits, the folded position
angles of the  outer companions are plotted  instead.  The correlation
between the angles is notable (correlation coefficient 0.58$\pm$0.13),
suggesting some alignment between the subsystems.  This plot uses only
the position angles,  and it can be produced for  a larger sample with
edge-on inner  orbits where  motions of the  outer companions  are not
measured yet and only their position angles are known.

The qualitative results of this section complement the analysis of
the sign correlation below from which the  triples with highly inclined orbits
and/or radial motions are excluded.

\subsection{Sign Correlation}
\label{sec:sign}

\begin{figure*}
\epsscale{1.1}
\plotone{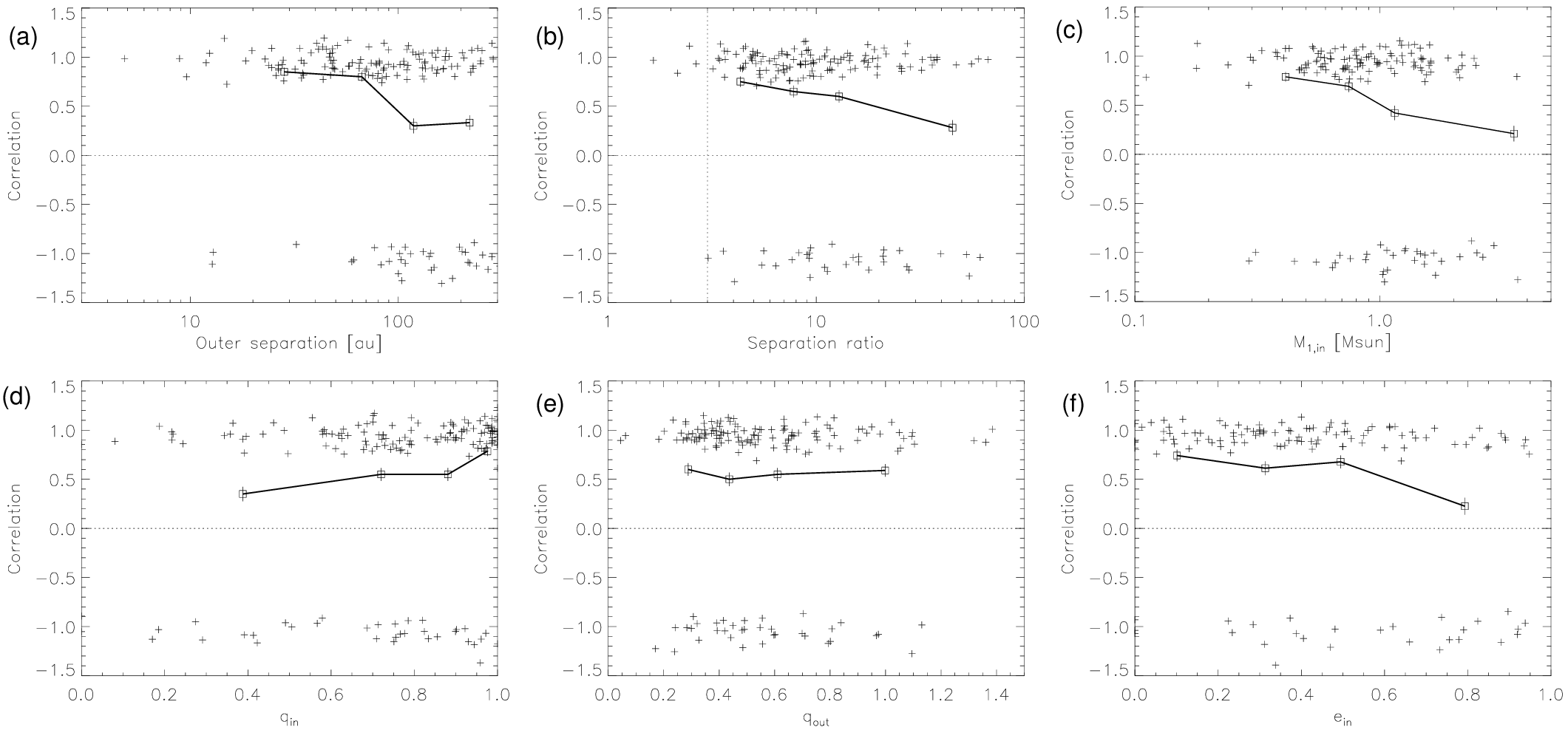}
\caption{Relative motion  direction $S$  vs.  several  parameters: (a)
  outer separation, (b) separation ratio,  (c) inner primary mass, (d)
  inner mass ratio, (e) outer mass ratio, and (f) inner eccentricity.
  The points are randomly displaced vertically to reduce overlap.  The
  sample  is split  by each  parameter  into four  equal groups.   The
  squares with  error bars and the  thick lines plot the  average sign
  correlation vs.  average parameter value in each group.
\label{fig:corr}
}
\end{figure*}

The motion  direction is known when  either orbits or linear  fits are
available.   To reduce  the impact  of errors,  I exclude  the edge-on
orbits with inclinations between 85\degr ~and 95\degr ~because without
complementary data on radial  velocities we cannot distinguish between
co- and counterrotating cases. For subsystems without computed orbits,
the motion direction is defined by the angle $\gamma$.  Accounting for
the measurement errors, the linear models  with $0 + 2 \sigma_\gamma <
\gamma < 180 - 2 \sigma_\gamma$ have direct motion, while for $180 + 2
\sigma_\gamma  <  \gamma  <  360  - 2  \sigma_\gamma$  the  motion  is
retrograde.  Filtering out the  edge-on orbits, noisy linear solutions
with  $\sigma_\gamma  >  20\degr$,   linear  models  with  unknown
  $\sigma_\gamma$,  and $\gamma$ directions  that are close to radial
within $2 \sigma_\gamma$ leaves 160 triples  for the study of the sign
correlation,  where  $S=+1$  stands  for  corotation  and  $S=-1$  for
opposite  rotation.      The  remaining  triples   have  $S=0$  in
  Table~\ref{tab:sample}.     The filtering  leaves only 57\%  of the
  original sample; however, the data for the full sample are provided,
  and a less restrictive analysis of  the sign correlation can be done
  by others.

Figure~\ref{fig:corr}  plots  $C =  \langle  S  \rangle$ vs.   several
parameters:  the outer  separation,  the separation  ratio, the  inner
primary  mass,  the  inner  and  outer  mass  ratios,  and  the  inner
eccentricity.   To avoid  overlap, the  points are  randomly displaced
vertically from $\pm$1.   To quantify the trends, the  sample is split
into four  equal  groups over each parameter, and the sign
correlation  $C$  in  each  group  is plotted  by  squares  and  lines
vs. average parameter in the group.   The errors are computed from the
binomial distribution, $\sigma_C = \sqrt{ (1 + C)(1 - C)/N}$, where $N
= 40$  is the group  size.  

The dependence of  orbit alignment on the outer separation  and on the
separation ratio, noted earlier  \citep{Sterzik2002}, is confirmed, as
well as the trend of decreasing alignment with increasing mass.  These
trends  are  not totally  independent  owing  to correlations  between
parameters; for  example, low-mass  triples tend  to have  larger mass
ratios.  The dependence  of the degree of alignment on  the inner mass
ratio is  a new  result.  On  the other  hand, there  is no  trend vs.
outer mass  ratio $q_{\rm  out}$.  Figure~\ref{fig:corr}f   plots
the sign correlation vs.  $e_{\rm in}$  for  125 triples with known
inner orbits and known outer motion direction.

Inner orbits  in corotating  triples are,  on average,  less eccentric
\citep{Tok2017}.  In the 98 triples with known inner orbits and $S=1$,
the median inner eccentricity is 0.37, while in the 27 counterrotating
triples  with $S  = -1$  it is  0.58.  The  actual difference  between
eccentricities is larger because $S$ has little meaning for individual
systems,  and among  the 98  systems with  $S=1$ there  are misaligned
ones.  However, statistically, the link between inner eccentricity and
alignment  is  established  reliably  (Figure~\ref{fig:corr}f).   Both
inner and outer  orbits in the aligned low-mass  triple system LHS1070
(Figure~\ref{fig:LHS1070}) are quasi-circular.   Although the trend is
clear,      there     are      notable     exceptions.      
In our  sample, there  are 30   inner  orbits with   eccentricity
above 0.7,  24 of which  have  known relative  motion.  The sign
correlation in this subset is low, 0.08$\pm$0.24.

For  the full  sample, $C  = 0.56  \pm 0.10$  corresponds to  the mean
mutual inclination  $\langle \Phi \rangle$ of  40\degr ~(this estimate
is approximate).   For the 65  triples with inner primaries  below 0.8
\msun, $C  = 0.75 \pm 0.16$,  while for the remaining  95 more massive
systems $C  = 0.43 \pm 0.12$  ($\langle \Phi \rangle$ of  23\degr ~and
59\degr,  respectively).   The  subsample  of 38  triples  with  inner
primaries below  1 \msun  and outer  separations below  50 au  has the
largest degree of  orbit alignment, $C = 0.89 \pm  0.22$ or $\langle
\Phi \rangle \approx 10\degr$.

The general  trends outlined here  are not absolute,  and hierarchical
systems contradicting  them are  found infrequently  but consistently,
like the  counterrotating low-mass double  twin in the lower  panel of
Figure~\ref{fig:motions}, or aligned low-mass triples with large inner
eccentricities.  For  example, in the triple  system 11247$-$6139 (HIP
55691)  with  a   0.74  \msun  inner  primary,   the  measured  mutual
inclination  between the  141  yr outer  and 2.4  yr  inner orbits  is
6\fdg5$\pm$0\fdg8.  Yet,  the inner eccentricity  is 0.8802$\pm$0.0003
\citep{chiron11}.  Another example is 08447$-$2126 (HIP 42910) with an
inner eccentricity  of 0.948$\pm$0.006  and apparently  aligned orbits
\citep{Trip2023}.

\subsection{Resolved Quadruple Systems}
\label{sec:quadr}

\begin{figure}
\epsscale{1.0}
\plotone{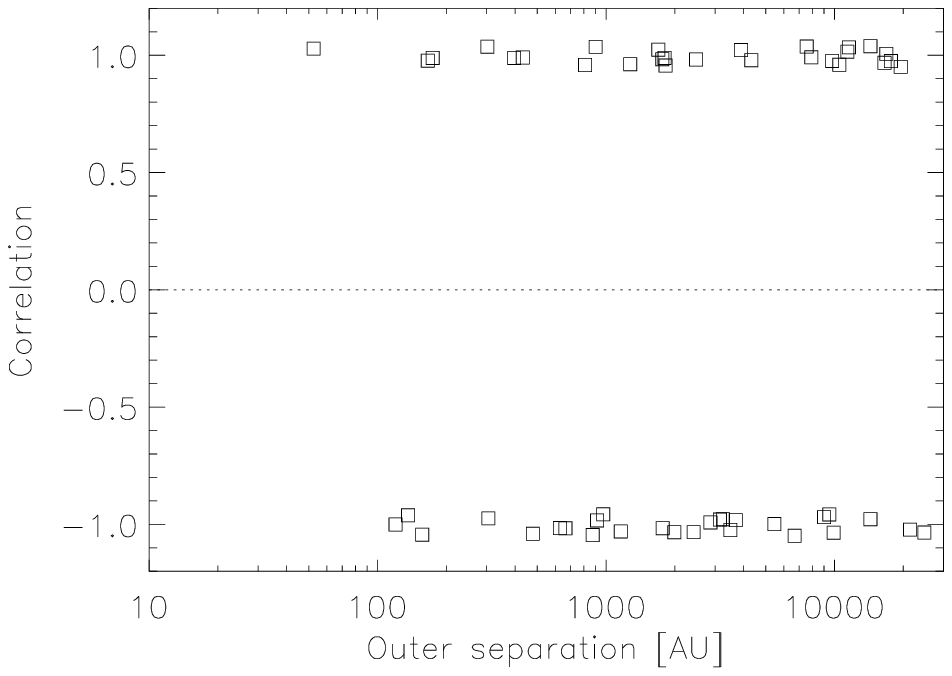}
\caption{Relative motion directions of two inner pairs in
  2+2 quadruples vs. outer separation.
\label{fig:quadr}
}
\end{figure}

Quadruple systems of 2+2 hierarchy where both inner pairs are resolved
are not  included in the main  sample.  However, it is  instructive to
take a look  at mutual orientations of inner orbits  in these systems.
Their list  based on the MSC  was examined manually to  identify cases
where  the  direction  of  motion  in both  inner  subsystems  can  be
determined either from  their orbits or, otherwise,  from the position
angle variation vs.  time (without fitting linear models).  In 63 such
systems the mean  sign correlation is $C = -0.14  \pm 0.12$.  However,
as  shown in  Figure~\ref{fig:quadr}, the  outer separations  in these
quadruples are larger  than in triples, with a median  of 2.9 kau and
only one below  100 au.  So, the  lack of  mutual alignment  between inner
orbits  in wide  2+2 quadruples  agrees  with the  statistics of  wide
triples.   Some  hierarchies in  this  2+2  sample contain  additional
subsystems, i.e.   more than  four components.  As  the 2+2  sample is
only illustrative, it is not tabulated  here.

Misalignment in  several relatively compact (outer  periods $<$500 yr)
2+2  quadruples where  inner  spectroscopic  subsystems were  resolved
interferometrically is  a rule  rather than  exception, for  example in
04357+1010        \citep[88       Tau,][]{Lane2007},        07346+3153
\citep[Castor,][]{Torres2022},      and     11221$-$2447      \citep[HD
  98800,][]{Zuniga-Fernandez2021}.  A misaligned  disk is found around
one of the inner pairs in HD 98800.

\subsection{Eclipsing Subsystems in Visual and Astrometric Pairs}
\label{sec:visecl}

\begin{figure}
\epsscale{1.0}
\plotone{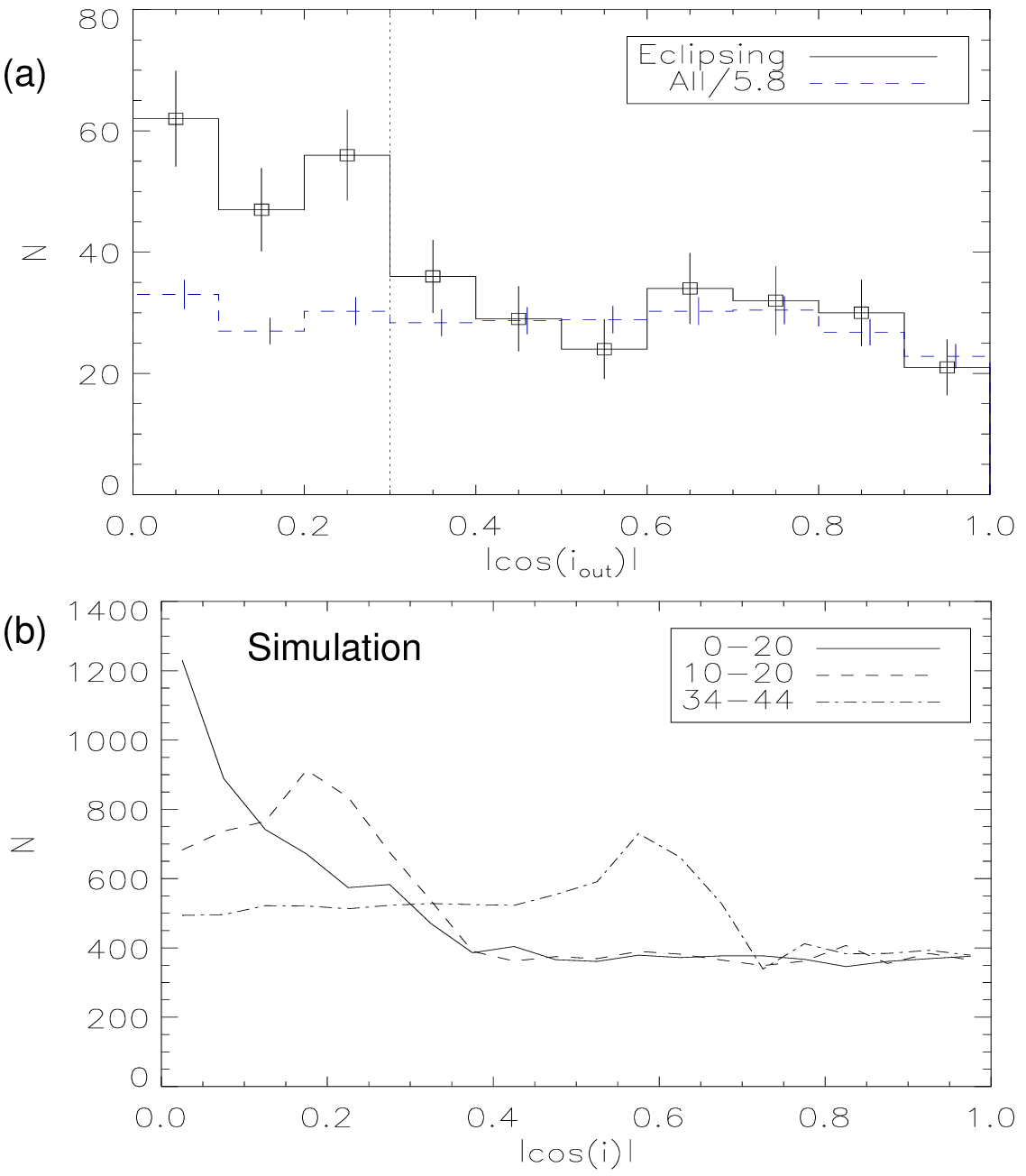}
\caption{The  histogram  of  $|  \cos  i_{\rm out}  |$  in  visual  or
  astrometric binaries  containing eclipsing subsystems is  plotted in
  panel (a) by  full line and squares  with error bars, and  the down-scaled
  histogram for all known orbital  inclinations is plotted by the blue
  dashed line.   Panel (b) shows  histograms for simulated  samples of
  triples where 0.75 fraction are  aligned randomly, and the rest have
  mutual inclinations  distributed in  three intervals: full  line ---
  (0,20),  dashed line  ---  (10,20), and  dash-dot  line ---  (34,44)
  degrees.
\label{fig:visecl}
}
\end{figure}

To  complement   the  study  of   resolved  triples,  I   explore  the
distribution of  known orbital inclinations in  pairs containing inner
eclipsing subsystems. Preliminary results of this study were presented
at a conference \citep{Kopal2024}.  As a  first step, a sample of 1657
subsystems with known visual or astrometric orbits is extracted
from the  MSC.  Most  of them are  relatively nearby  (median parallax
11.7\,mas). A  subset of  this sample  containing 371  inner eclipsing
binaries  is  defined.   The  majority of  these  triples  are  recent
discoveries  resulting from  matching catalogs  of eclipsing  binaries
with the  Gaia catalog  of astrometric  orbits \citep{Czavalinga2023};
their outer periods  are mostly shorter than 1000 days  (the time span
of the GDR3 mission).  On  average, triples containing eclipsing pairs
(called VA+E for  brevity) are more distant, their  median parallax is
2.3\,mas.  The  median masses in  the full  orbital sample and  in its
VA+E subset are 1.19 and 1.37 \msun, respectively.

\begin{deluxetable}{l   c  c c }     
\tabletypesize{\scriptsize}  
\tablecaption{Alignment in the VA+E Sample
\label{tab:visecl} }   
\tablewidth{0pt}   
\tablehead{        
\colhead{Sample} &
\colhead{$N$} &
\colhead{$N_{\rm align}$} &
\colhead{$f_a$} 
}
\startdata
All                  & 371 & 80.4    & 0.217$\pm$0.040 \\
$P_{\rm out} < 1000$ d & 258 & 72.2    & 0.280$\pm$0.049 \\ 
$P_{\rm out} > 1000$ d & 113 &  8.2    & 0.073$\pm$0.068 \\
$e_{\rm out} < 0.4$    & 203 & 54.6    & 0.269$\pm$0.055 \\
$e_{\rm out} > 0.4$    & 168 & 25.8    & 0.154$\pm$0.057 \\
$P_{\rm in} < 1$d         & 197 & 38.8    & 0.197$\pm$0.054 \\
$P_{\rm in} > 1$d         & 174 & 41.6    & 0.239$\pm$0.058 \\
\enddata 
\end{deluxetable}

The inclination  of eclipsing subsystems  should be close  to 90\degr,
and  if they  are aligned,  a preference  of edge-on  outer orbits  is
expected.  For orbits oriented randomly  with respect to the observer,
the cosine of  inclination is distributed uniformly  in $[-1,1]$.  Any
alignment should  be manifested by  an excess of edge-on  outer orbits
with $|\cos i_{\rm out} | \sim  0$.  The histogram of this quantity is
shown in the upper panel of Figure~\ref{fig:visecl}.  The distribution
for  the full  sample with  orbits (blue  dashed line)  appears almost
uniform, but  in the VA+E subsample  there is an excess  between 0 and
0.3.

To quantify the degree of alignment, I use the numbers $N_1$ and $N_2$
of the VA+E systems  with $| \cos i_{\rm out} | $  below 0.3 and above
0.5,  respectively.  Then  $N_{\rm align}  =  N_1 -  (3/5)N_2$ is  the
excess of  aligned systems relative  to the uniform  distribution. The
fraction  of this  excess in  the  full sample  $N$ is  $f_a =  N_{\rm
  align}/N$. Its error is estimated  approximately by the Poisson law,
$\sigma_f = \sqrt{ N_1  + 0.6^2 N_2}/N$.  Table~\ref{tab:visecl} lists
these parameters for the full sample and  its cuts.

The first  conclusion is that  about 78\%  of hierarchies in  the VA+E
sample do not show any alignment between their inner and outer orbits,
while the  remaining 22$\pm$4\% are aligned  within $| \cos i_{\rm  out} | <
0.3$ ($i_{\rm  out}$ of  90\degr$\pm$17\fdg5).  The excess  of aligned
subsystems  is statistically  significant at  the $5.4  \sigma$ level.
Note  that the  orbits  of eclipsing  binaries  may have  inclinations
different  from 90\degr,  while the  inclinations  of visual  and
astrometric  orbits are  affected  by the  measurement errors.   These
effects broaden  the observed distribution,  so its width is  an upper
limit of the true mutual alignment.

The  fraction of  aligned hierarchies  in the  VA+E sample  apparently
depends on  the outer  period, being larger  for compact  triples with
$P_{\rm out} < 1000$  days.  Table~\ref{tab:visecl} hints that triples
with less eccentric  outer orbits or longer inner  periods are aligned
more  frequently,  although  the  differences  are  not  statistically
significant.  No  difference in  alignment is  found by  splitting the
sample around 1.3 \msun mass.

The width of  the cosine distribution in the  aligned systems suggests
mutual inclinations $\Phi$ within  $\sim$20\degr.  To quantify better,
I  simulated  $10^4$   triples  where  a  0.25   fraction  has  mutual
inclinations  $\Phi$  uniformly  distributed in  specified  intervals,
while  the   rest  have   random  alignment.    The  lower   panel  of
Figure~\ref{fig:visecl}    shows   three    representative   simulated
distributions  of   $|  \cos  i_{\rm   out}  |$.   A   uniform  $\Phi$
distribution between 0\degr  ~and 20\degr ~produces a  sharp peak near
zero that does not resemble  the actual histogram, and a concentration
of   $\Phi$   around  39\degr   predicted   by 
\citet{Fabrycky2007}  is also  excluded.  A  qualitative match  to the
observed  histogram shape  is found  for $\Phi$  between 10\degr  ~and
20\degr.  However, these simulations do  not account for the histogram
broadening by measurement errors.

In the majority  of misaligned compact triples, the  precession of the
inner  pair  should  change  the inclination  and,  consequently,  the
eclipse depth.  However, typical  precession periods are long ($\sim$3
kyr for inner and outer periods of 1 and 1000 days), leaving detection
of such effects for the future, except the few known examples.

\section{Discussion}
\label{sec:disc}

\subsection{Theoretical Predictions}
\label{sec:theory}

\begin{figure*}
\epsscale{0.7}
\plotone{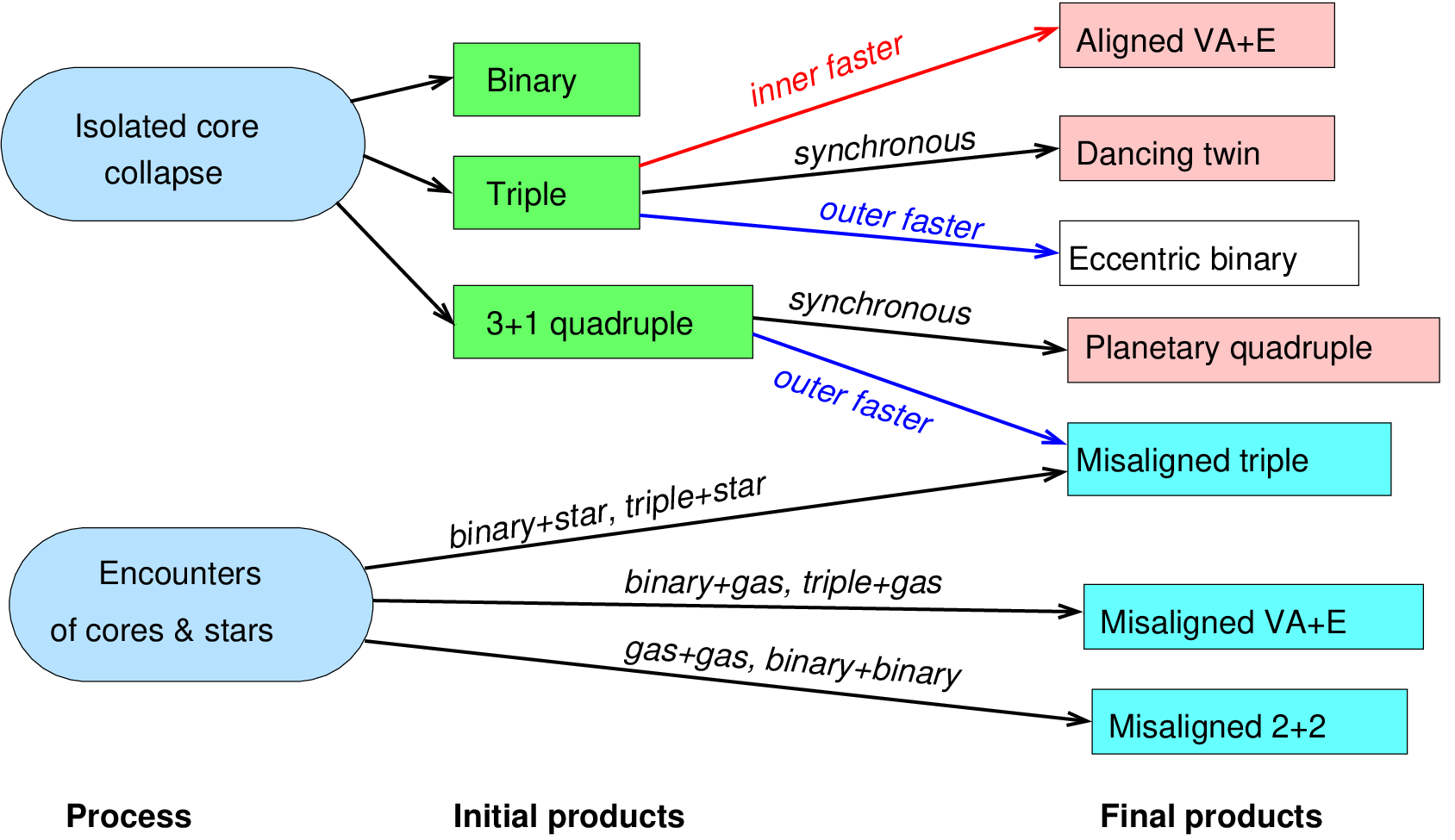}
\caption{Formation  and   early  evolution  of   hierarchical  systems
  (section~\ref{sec:theory}).   VA+E stands  for compact  triples with
  eclipsing inner pairs. 
\label{fig:evolution}
}
\end{figure*}

To put this  work in proper context, a broad  outline of the formation
of stellar hierarchies is  in order. Figure~\ref{fig:evolution} offers
a visual guide  to the following text;  various processes contributing
to   the   architecture  of   multiple   systems   are  discussed   in
\citet{Mult2021}.

Stars form in groups: clusters, associations, and small aggregates.
Nascent  groups   are  highly   structured,  inheriting   the  spatial
distribution of the molecular gas,  and this reflects in the hierarchy
of young multiple systems at large spatial scales \citep{Joncour2017}.
Interactions between the stars are  governed by the Newtonian dynamics
of point masses (N-body). The decay of young stellar groups inevitably
leaves some  surviving bound  wide triples.  Their  architecture bears
imprints of the chaotic dynamics: the orbits are not aligned mutually,
the inner  orbits have the  thermal eccentricity distribution  $f(e) =
2e$, and the ratios of outer and inner separations are modest, not too
far above  the dynamical  stability limit  \citep{Antognini2016}.  The
properties  of wide  triples  in  the field    match the  N-body
predictions  \citep{Gaiatriples,Shariat2025b}.    However,  they  show
signs of weak orbit alignment at outer separations below $\sim$100 au,
where the dynamical effect of the gas becomes important.

Median separations of  binaries are in the  range of 10 to  100 au, so
the architecture of  typical stellar systems depends  on the processes
involving  gas,  such  as   core  fragmentation,  disk  fragmentation,
gas-assisted     capture,     and    accretion     \citep{Offner2023}.
\citet{Rohde2021}  explored numerically  the  collapse  of isolated  1
\msun cores  of $\sim$3 kau  initial radius; on average,  they produce
3.1 stars  per core. The  mass ratios  $q$ of these  low-mass binaries
mostly  exceed  0.6,  and  43\%  are twins  with  $q  >  0.95$;  their
separations are below  100 au.  The outflows increase  the fraction of
twins by blocking  gas supply from polar directions  and enhancing 
infall from the  equatorial plane, where most mass is  accreted by the
secondary component.   Although some triples are  dynamically unstable
and decay,  new systems form,  so the fraction of  hierarchies remains
almost  constant  at  0.17$\pm$0.03  during the  simulations.   It  is
substantially  higher  than  $\sim$0.05  for stars  of  0.5  \msun  in
Figure~1 of  \citet{Offner2023}.  The  relatively large  percentage of
hierarchies produced in the  simulations of \citet{Rohde2021} could be
related to  the core  rotation.  Unfortunately,  their paper  does not
explore the orbit alignment in triples.  \citet{Bate2014} presents the
multiplicity  statistics resulting  from  his  simulations of  cluster
collapse.  The orbits  of 31 triples are relatively  well aligned with
$\langle \Phi \rangle = 39^\circ$.  A dependence of the alignment on the
outer  separation and  period ratio  is found  (see their  Figure 20),
matching the observed trends.

The accretion of gas accompanied  by inward migration is the generally
accepted process  creating close binaries, including  inner subsystems
in triples.  The  partial alignment of triples  with outer separations
below  100 au  is presumably  caused by  the accretion  of gas  with a
relatively constant  angular momentum inherited from  the parent core;
it  naturally  extends to  the  alignment  between binary  orbits  and
stellar rotation  axes \citep{Bate2014}.  The aligned  gas is accreted
preferentially by  the secondary component, increasing  the mass ratio
and producing twins  \citep{TokMoe2020}.  Many twins are  found in the
inner subsystems of hierarchies, and double twins where both inner and
outer  mass ratios  are close  to one  are also  typical for  low-mass
hierarchies  with  moderate  outer  separations  \citep{Dancingtwins},
including   in this sample. Such  hierarchies, ``dancing twins'',
are also approximately aligned and have moderately eccentric orbits. A
related class  are the  aligned 3+1  ``planetary'' quadruples  such as
HD~91962 \citep{HD91962}  or the compact  (periods 2.9 days,  59 days,
and 3.9 years) low-mass quadruple HIP~41431 \citep{Borkovits2019}.

Although  approximately  aligned triples  and  3+1  quadruples can  be
linked  to  the  products   of  isolated  core  collapse,  theoretical
predictions of their properties are still vague.  If the infalling gas
is accreted mostly by the outer  component of a triple and it migrates
faster  than  the inner  binary,  the  system may  become  dynamically
unstable  and  disrupt,  leaving  behind an  eccentric  binary  (or  a
misaligned triple after decay of a quadruple).  On the other hand, the
migration  in   the  inner  and   outer  subsystems  can   be  somehow
synchronized  at  the period  ratio  on  the  order  of 20,  when  the
subsystems already  interact mutually  but the triple  remains stable;
such period  ratios are typical  for dancing twins  and planetary-type
quadruples.   The existence  of  compact (e.g.   eclipsing and  triply
eclipsing)  and aligned  hierarchies \citep{Borkovits2016}  indicates
that  migration  can  shrink  inner orbits  very  efficiently  without
destroying the system.   The three possible types  of accreting triple
evolution (inner faster, synchronized, and  outer faster) are coded by
the    red,    black,    and     blue    lines,    respectively,    in
Figure~\ref{fig:evolution}.

Interactions  of the  products of  core fragmenation  with neighboring
stars and gas  (e.g. between two fragments of the  filament falling on
to the common  center of mass) can alter the  final stellar systems or
trigger the collapse.   \citet{Whitworth2001} proposed that collisions
between two cores  can spawn relatively wide 2+2  quadruples. A triple
formed  by an  isolated  core  collapse can  be  disrupted by  another
approaching  star; an  encounter between  two binaries  can produce  a
misaligned triple.   A misaligned gas  falling on  to a triple  can be
captured  preferentially  by the  inner  subsystem,  causing its  fast
migration and  producing a  triple with  misaligned tight  inner pair.
This  could be  a dominant  channel of  close binary  formation within
triples.    Encounters  could   potentially   explain  the   empirical
correlation between  close binaries and hierarchical  multiplicity and
the correlated  occurrence of  subsystems in  both components  of wide
binaries   inferred from  the relatively  high frequency  of 2+2
quadruples in  the field \citep{FG67b}.  However,  theoretical studies
of  these mechanisms  yielding  quantitative  predictions are  needed.
Misaligned gas  structures are directly observed  in young hierarchies
such as  the quadruple  HD 98800 \citep{Zuniga-Fernandez2021}  and the
triple GW~Ori,  a member of our sample \citep{Kraus2020}.

The multiplicity is a strong function of mass \citep{DK13,Offner2023}.
Formation of massive stars and  stellar systems requires mass assembly
from   a   large   volume   \citep{Clark2021,Hennebelle2024}.    Their
properties are shaped  by interactions with stellar  neighbors and gas
outside the  original cores to  a larger extent, compared  to low-mass
hierarchies.   Qualitatively, the  dependence of  the architecture  of
stellar systems on  their mass and size agrees with  the expected role
of  encounters.  The   observed  trends can  be interpreted  as a
mixture in  varying proportion between approximately  aligned pristine
hierarchies  and misaligned  products of  their interaction  with the
environment.

\subsection{Inferences from This Study} 

The sample of  resolved triples with outer  separation below $\sim$300
au reveals clear trends in the mutual orbit alignment and mass ratios,
 in agreement with previous studies.  These trends can be matched
qualitatively  to the  mechanisms  of  triple-star formation  outlined
above.

\begin{enumerate}
\item
Alignment decreases  with increasing outer separation  and vanishes at
$s >  300$ au  because wider  triples are  shaped by  N-body dynamics,
while gas  dynamics is  relevant at spatial  scales comparable  to the
size  of accretion  disks  and  smaller.  The  influence  of gas 
  accretion is also  demonstrated by the substantial fraction of
inner twins and  by moderate inner eccentricities  in aligned triples,
 which  were presumably  damped by dissipation  in a  disk.  
  Most triples with  eccentric inner  orbits ($e_{\rm in}  > 0.7$)  are not
  aligned.

\item
The  alignment decreases  with  increasing ratio  of  outer and  inner
separations or  periods, with  decreasing inner  mass ratio,  and with
increasing mass.   These trends can  be explained by the  accretion of
misaligned gas unrelated to the original core.  The average mutual
  inclination in  the full sample  is 40\degr, but  in a subset  of 38
  triples with primary components less  massive than 1 \msun and outer
  separations below 50 au it is 10\degr.

\item
Triple systems  in non-hierarchical configurations (``trapezia'')
are found  not only at  wide separations, but  also at orbit  sizes of
$\sim$10 au. These field triples  are close to the dynamical stability
limit.  Their existence suggests  that migration could destabilize and
disrupt  relatively compact  triples,  and that  low-mass trapezia  in the
field (Figure~\ref{fig:LHS1070}) are their surviving subset.

\item
A substantial  fraction of triple  systems in  this sample (46  out of
 278)  have additional  distant companions  at separations  of $10^3  -
10^4$  au.    This  indicates   that  they   were  formed   in  sparse
environments, rather  than in clusters.  Their  architecture therefore
is relevant to fragmentation of isolated cores.

\item
 Inner  orbits  in 63  2+2   quadruples  with  outer  separations  
from 100 to $10^4$ au are not mutually aligned.

\item
 Only 22\% of triples with known outer orbits and inner eclipising
  subsystems are  mutually aligned within $\sim$20\degr,  the rest are
  aligned  randomly. The  fraction of  aligned triples  is larger  for
  outer periods under 1000 days. 

\end{enumerate}

This study explores the separation regime from a few to 300 au, on the
order  of median  separation of  low-mass binaries.   Its results  are
complemented  by  the statistics  at  larger  separations provided  by
\citet{Shariat2025b} and  \citet{Gaiatriples}, showing  the transition
between close (gas-dominated) and  wide (N-body) regimes.  Apparently,
collapse  of  isolated cores  typically  produces  low-mass stars  and
well-organized planar  hierarchies.  At still smaller  separations, we
expect  an even  stronger imprint  of  the gas  accretion, leading  to
aligned orbits.   The compact eclipsing triples  have a preferentially
flat  architecture  and  are  often   aligned  within  a  few  degrees
\citep{Borkovits2016,Borkovits2022};  for   this  reason,   many  such
hierarchies are triply eclipsing. However, the properties of eclipsing
subsystems explored in section~\ref{sec:visecl}  are at odds with this
view, showing that  in 3/4 of such hierarchies, even  in compact ones,
the relative  orbit alignment  is random.   The division  of eclipsing
triples  in two  groups (aligned  and  random) and  the outliers  
  contradicting the general trends suggest that hierarchical systems were
formed    via    different    processes   and    their    combinations
(Figure~\ref{fig:evolution}).

\subsection{Future Work}
\label{sec:future}

The complexity of triple-star  formation reflected in their statistics
tells  us  that  considering  these  objects  without  distinction  in
separation, mass, etc.,  as a single population, hides  the trends and
is not very informative. Early  statistical studies did not allow such
distinction owing  to the small samples.   Ground-based monitoring and
space surveys will enlarge the samples  to clarify known trends and to
find new ones. The SOAR program on multiple stars illustrates the need
of  a long  time coverage  combined with  high angular  resolution for
revealing the  architecture of stellar hierarchies.   Its continuation
and extension to the northern sky would be promising.

The diversity of initial conditions and processes involved in the star
formation stand  on the  way of  predicting statistical  properties of
stellar systems.   While a general  understanding is reached,  and the
simulations match  the observed  statistics qualitatively,  much work
remains   to   be    done.    Large-scale   hydrodynamic   simulations
\citep{Bate2014}  are  expensive and  have  a  low statistical  yield.
Targeted simulations  on smaller scales  sampling a range  of physical
parameters, initial conditions, and random realizations, like those of
\citet{Rohde2021}, is a promising way  forward.  The role of accretion
in shaping the masses and orbital architecture of binaries and triples
is not yet fully explored,  despite several studies.  For example, the
synopsis  of   hydrodynamic  simulations  of  accreting   binaries  by
\citet{Valli2024} treats  only the planar two-dimensional  case, while
in  reality  the  misalignment of gas  is   essential,  and  the  problem  is
intrinsically three-dimensional \citep{Smallwood2025}.

\begin{acknowledgments}

The  research  was funded  by  the  NSF's  NOIRLab.  Comments  by  the
anonymous Referee helped to improve the clarity of presentation.  This
work  used  the  SIMBAD  service  operated  by  Centre  des  Donn\'ees
Stellaires  (Strasbourg, France),  bibliographic  references from  the
Astrophysics Data  System maintained  by SAO/NASA, and  the Washington
Double Star  Catalog maintained at  USNO.  This  work has made  use of
data   from   the   European   Space   Agency   (ESA)   mission   Gaia
(\url{https://www.cosmos.esa.int/gaia}),  processed by  the Gaia  Data
Processing        and         Analysis        Consortium        (DPAC,
\url{https://www.cosmos.esa.int/web/gaia/dpac/consortium}).    Funding
for the DPAC has been provided by national institutions, in particular
the institutions participating in the Gaia Multilateral Agreement.

\end{acknowledgments}






\facility{SOAR, Gaia}

\bibliography{triples.bib}{}
\bibliographystyle{aasjournal}

\end{document}